# Simple Single-Section Diode Frequency Combs

Matthew. W. Day[1], Mark Dong[1], Bradley C. Smith[1], Rachel C. Owen[1], Grace C. Kerber[1], Taigao Ma[1], Herbert G. Winful[1,2], and Steven T. Cundiff[1,2] [*]

[1]Department of Physics, University of Michigan, 450 Church Street, Ann Arbor, MI 48109-1040
[2]Department of Electrical Engineering and Computer Science, University of Michigan, 1301 Beal Avenue, Ann Arbor, MI 48109-2122
[*]cundiff@umich.edu

**Abstract**

Frequency combs, broadband light sources whose spectra consist of coherent, discrete modes, have become essential in many fields. Miniaturizing frequency combs would be a significant advance in these fields, enabling the deployment of frequency-comb based devices for diverse measurement and spectroscopy applications. We demonstrate diode-laser based frequency comb generators. These laser diodes are simple, electrically pumped, inexpensive and readily manufactured. Each chip contains several dozen diode-laser combs. We measure the time-domain output of a diode frequency comb to reveal the underlying frequency dynamics responsible for the comb spectrum, conduct dual comb spectroscopy of a molecular gas with two devices on the same chip, and demonstrate that these combs can be battery powered.

Optical frequency combs[1,2] initially attracted attention for their impact on the field of precision measurement where they revolutionized optical frequency metrology[3,4] and synthesis[5] and enabled optical atomic clocks.[6] These applications leverage the ability of combs to coherently link disparate optical frequencies and directly connect optical and radio frequencies. A second wave of interest was triggered by the development of dual-comb spectroscopy and related methods,[7-11] which rely on the mutual coherence of two combs to realize optical spectroscopy with a combination of spectral resolution, acquisition time and size that cannot be achieved using conventional methods. The development of methods to measure and control the frequency comb generated by mode-locked lasers[1,12] was key to these applications.

In parallel with the broadening of applications based on highly controlled combs, new sources of combs were being developed. One prominent current approach is based on nonlinearities in micro-resonators.[13] This approach features a chip-scale resonator with large free-spectral range, although the required high power continuous-wave pump presented a challenge to realizing fully chip-scale systems, which has only recently been achieved.[5,14,15] Another important approach was launched by the discovery that quantum cascade lasers (QCLs) could produce frequency combs.[16] While QCLs are compact and electrically pumped, their development was important as they provide a simple source for mid-infrared (IR) to terahertz frequency combs, although the long operating wavelengths brought with them a need for cryogenic cooling.

Directly generating combs from inter-band diode lasers in the near-IR or visible is certainly attractive as they are electrically pumped, compact and do not require cryogenic cooling. While there have been scattered reports of comb generation in single section devices,[17,18] most effort on engineering diode lasers to produce combs has focused on traditional mode-locking, which is designed to produce a train of ultrashort pulses. However, the deleterious gain dynamics in semiconductors have made this challenging.[19] Consequently, diode-laser combs have lagged behind other sources, for example dual comb radio frequency (RF) spectra have only recently been demonstrated.[20,21] Success in developing an inexpensive, miniature, portable, simple and low power comb source, whether a diode laser or otherwise, would greatly extend the reach of dual-comb spectroscopy to provide a path to ubiquitous spectroscopy for a large variety of applications such as atmospheric monitoring, pollution monitoring, chemical sensing or breath analysis. One could even imagine them being incorporated into mobile phones, as accelerometers are today.

To this end, we develop and characterize the electric field output of electrically pumped, single section diode lasers outputting frequency combs.[22] These devices are monolithic and cheap to manufacture. Each diode's volume is ~0.01 mm$^3$ allowing several dozen devices fit on a 0.5 × 2 × 1 mm chip. We demonstrate tuning of individual comb teeth over multiple free-spectral ranges (FSR) by adjusting pump temperature or current, fulfilling a key requirement for combs used in precision metrology. Furthermore, we demonstrate dual comb spectroscopy of molecular transitions,[20] and power our chip scale combs with AA batteries. These results show that diode frequency combs (DFCs) have sufficient coherence to be used in dual comb spectroscopy and thus realize the vision for ubiquitous and portable comb-based spectroscopy. The epitaxial growth of our diode laser layers was performed by a commercial foundry, enhancing the possibility for them to be manufactured inexpensively, using materials standardized the by telecommunications industry, an important distinction from the recent demonstration of dual comb spectroscopy at 2 microns.[21]

Microscopic ring resonators are the current state of the art in near-IR compact frequency comb sources. The resonators are typically rings of silicon nitride or fused silica pumped externally by high-power erbium-doped fiber amplifiers, although multi-section single-comb, battery powered sources have recently been developed.[14,15] Entering the most useful low-noise single-soliton comb states requires a complex algorithm of parametric pump detuning.[23,24]

An alternative route has been the development of miniaturized semiconductor comb sources. Due to their broader gain bandwidth through inhomogeneous broadening, quantum dash frequency combs[25,26] and quantum-dot frequency combs[27,28] are more common than quantum well based devices.[17,18,20,29] Quantum well devices, however, have the advantages of higher gain coefficients and ease of production. Furthermore, due to the higher gain coefficients, quantum-well devices are more flexible because the repetition rate can be easily tuned over a wide range (~10-100 GHz) by varying quantum well number and device length, and offer a path toward convenient and compact comb generation in the visible and near-infrared spectral regions.[30] Furthermore, the gain bandwidth of quantum-well devices can be increased by using multiple quantum wells with varying width.[20]

In a coherent optical frequency comb, the frequency of every output mode is determined by two radio frequencies: the repetition rate ($f_{rep}$), or the inverse of the cavity round-trip time, and the offset frequency ($f_o$), which is set by the dispersion of the cavity. The frequency of the $n^{th}$ mode of a comb's output is $f_n = nf_{rep} + f_o$. This relationship undergirds the fields of precision optical frequency measurement and generation.

Two key properties are required for a laser to generate a coherent frequency comb: consistent gain for multiple modes simultaneously, and a mechanism to force coherence between adjacent cavity modes. For DFCs and other semiconductor devices that rely on frequency-modulated (FM) mode locking,[22] spatial hole burning (SHB) supports consistent gain between multiple modes, satisfying the first requirement. We ensure the presence of SHB by choosing the operating wavelength to be shorter than the ambipolar diffusion length of carriers[20] because each standing-wave cavity-mode 'burns' a spatial hole in the gain distribution, but gain will still be available between field maxima. In order to allow comparable gain amongst multiple modes, and ensure *coherent* output, four-wave-mixing mediated by the Kerr effect between modes enforces that the phase of each mode satisfies $\Phi = 2\psi_0 - \psi_+ - \psi_- = \pm\pi, 0$ where $\psi_0$ is the phase of an arbitrary lasing mode, and $\psi_\pm$ are the phases of the adjacent modes.[22] The solution $\Phi = 0$ (corresponding to a pulse train) requires a saturable absorber, but the solutions $\Phi = \pm\pi$ force the electric fields out of phase, maintaining the SHB-induced gain profile. The net result is that DFCs output a frequency- and amplitude-modulated comb.

Frequency combs are generated in p-n InGaAsP laser diodes with four quantum wells providing gain.[20] The active regions are composed of a quaternary InGaAsP quantum well structure, with a ridge waveguide providing lateral mode confinement. Device construction details can be found in the Supplement and reference 20. Chips, shown in Fig. 1(a), are typically 1-2 mm long and 2-4 mm wide, containing up to two dozen diodes. DFC comb generation, shown in Fig. 1(b), initiates once the pump current is above lasing threshold. The laser output spectrum stabilizes only when comb generation commences (typically observed for currents ~10-20% above the lasing threshold). The diodes were typically operated between 10 °C and 23 °C, with injection currents of between 110 mA and 230 mA (although currents >300mA were achieved). The diodes had a series resistance of ~1.5 Ω, leading to a power dissipation of ~0.3 W at an injection current of ~200 mA. Tungsten probes contacted the diode's top face, and a brass plate was used as a current return from the bottom face.

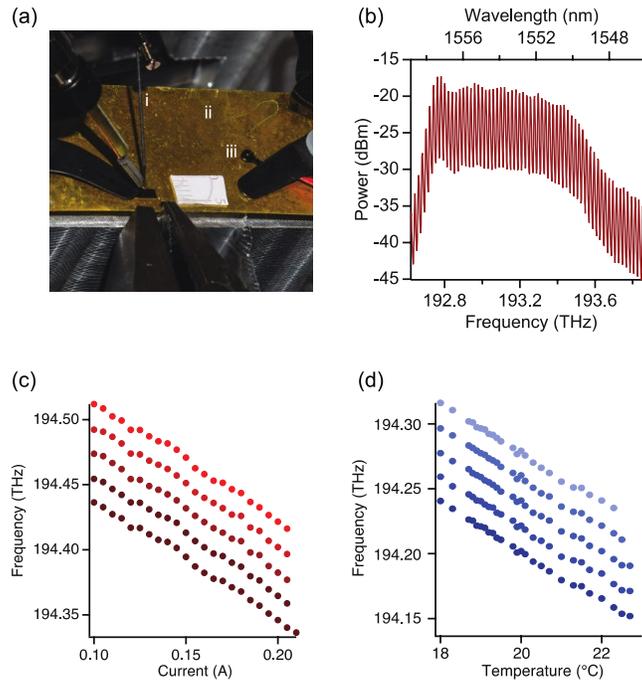

**FIG. 1** (a) a picture of a DFC chip containing 21 diodes: (i) is a tungsten probe, (ii) is the return electrode, under which a thermo-electric cooler sits and (iii) is a thermistor. A 5 mm ruler is present for scale. (b) A representative output spectrum of a DFC device. The center frequencies of a subset of comb teeth tracked as (c) pump current and (d) temperature were increased.

Operating wavelength, power, and repetition rate are coarsely set by device design and fabrication. A typical spectral width (-5 dB from the peak) is 0.8 THz (6 nm) at 1550 nm, while the repetition rate (roughly set by the cavity length) was in the range of 19-25 GHz. Fine control of $f_{rep}$ and $f_o$ is achieved by adjusting pump current and diode temperature, tuning individual comb teeth over multiple FSR as seen in Fig. 1(c) and (d). The ability to tune over multiple FSR distinguishes these devices from micro-resonator-based comb generation. Typical tooth widths were on the order of 200 kHz. The output spectrum can be broadened by the addition of quantum wells of other widths,[20] while the output power (typically 1mW) can be increased by adding quantum wells.

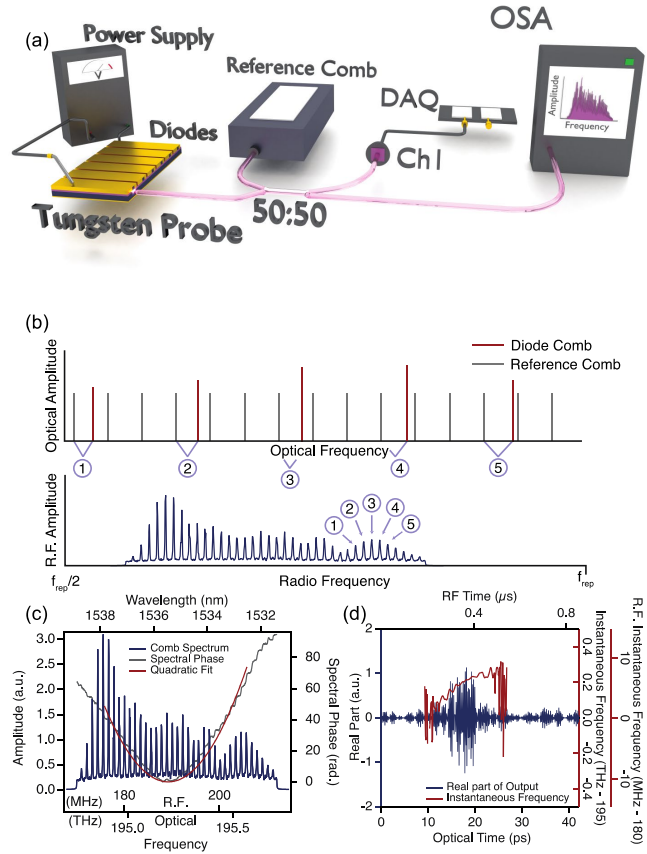

**FIG. 2** (a) Light from a DFC is combined with a reference comb on a 50:50 beam splitter, with one port directed to a fast photodiode and the other to an OSA. (b) a frequency domain cartoon of the XCDC measurement. The reference comb provides 100 teeth between each DFC comb tooth. (c) The extracted frequency domain output of the DFC device. (d) The real part and instantaneous frequency of the DFC time-domain output field.

The basic property upon which all frequency comb applications rely is mutual coherence between the comb teeth. Mutual coherence is typically difficult to prove and requires measuring the spectral phase profile of the comb spectrum. It is not enough to simply show a comb-like power spectrum to confirm that a light source is a coherent frequency comb. To show that the DFC output fulfills this requirement, we performed a cross-correlation dual comb (XCDC) measurement, shown in Fig. 2(a) and (b). This technique derives from traditional dual comb techniques[7,31] but is broadly applicable to rapidly characterizing the electric field output of frequency combs. We use it to confirm the prediction that the observed combs are due to frequency modulation mode-locking, as occurs in QCLs,[16,22,32,33] and has been hypothesized for diode lasers.[22] In addition, the XCDC measurement confirms the overall phase coherence of the DFC output.

To realize XCDC, we combined the DFC output ($f_{rep}$ = 25.0012 GHz) with that of a dispersion-compensated frequency comb with $f_{rep,r}$ = 250 MHz. In the time domain, this corresponds to a cross-correlation between the pulse trains of the DFC and reference comb. In the radio frequency domain, equivalent image spectra occur every $f_{rep,r}/2$. We acquired 350,000 samples at 500 MS/s, taking the data from 125 MHz to 250 MHz to be the XCDC signal (the frequency interval from $f_{rep,r}/2$ to $f_{rep,r}$ was chosen for technical reasons). Since DFCs were stable over seconds, we converted between the RF and optical domains by fitting the RF spectra to traces taken on an optical spectrum analyzer (OSA). After coherent averaging of the time-domain data to remove the effect of crosstalk in the acquisition electronics (see Supplemental Material), we unwrapped the spectral phase in the frequency domain by monitoring point-to-point changes in the spectral phase and adding the appropriate value of either $\pm 2\pi$ when the discontinuity exceeded $\pm \pi$ radians. As seen by the good qualitative fit of the data in Fig. 2(c), the spectral phase was parabolic in good agreement with theory presented in references 22 and 30. This fit allowed us to extract the estimated GDD of -4.3 ps$^2$, which corresponds to the lowest frequency portion of the comb arriving at the detector first. For details on the phase convention used in our analysis, see the Methods and Materials section. Device output [Fig. 2(d)] is periodic, with strong amplitude and frequency modulated components. We note that the reconstructed quasi-linear chirp in Fig. 2(d) is similar to that observed from quantum cascade lasers.[33]

To establish mutual phase coherence between two comb sources, it is not enough to simply demonstrate a comb structure in their RF beat spectrum. One must also demonstrate that the comb teeth in the spectrum are coherent with each other. This requirement is more challenging and requires that the phase noise in the RF spectra be correctible. To demonstrate this, we independently powered two DFCs on the same chip with injection currents of ~210 mA and ~195mA, and an operating platform temperature 13.3 °C. We then combined their outputs and measured RF dual comb spectra (DCS) before and after a 300 Torr H$^{13}$C$^{14}$N calibration cell along with an attendant OSA trace.[7,34-36] This high-pressure calibration cell was chosen because it has transition linewidths of roughly ~25GHz, well-matched to the repetition rate of our devices. Figure 3(a) presents a schematic of this experiment. The DCS data was collected over 10 µs at 2 GS/s and noise-corrected (see the supplemental information for details).[37-39] This process produced a pair of well-corrected RF combs, from which the amplitude of each comb tooth was used to calculate the absorption profile using Beer's law, comparing the result to a typical spectrum taken using a broad-band light source. Figure 3(c) shows a typical DCS absorption measurement. This measurement is similar to that in references 34 using a mode-locked laser and 35 using a microresonator. It demonstrates the ease of taking useful multi-comb spectra with even free-running, miniaturized DFC devices on the same chip at ambient operating conditions. Furthermore, due to the nature of DFCs, the repetition rate can be chosen during manufacture to be optimally matched to remote spectroscopy applications. Figure 3(c) shows the spectrum of a DFC device powered by AA batteries (two batteries provide roughly an hour's operation at present efficiency).

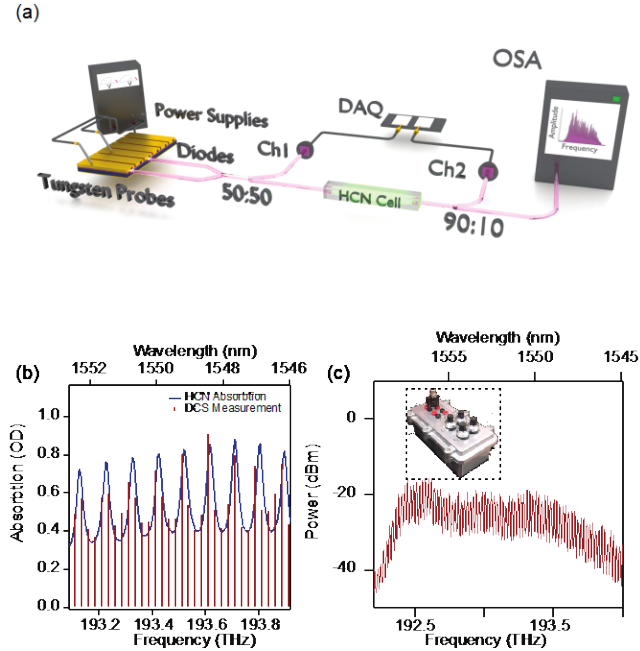

**FIG. 3** (a) The DCS experimental setup: independent power supplies provide current to two DFC devices. The light emitted from the devices is collected using tapered fibers and combined. Half of the resulting light is sent to a photodiode before the cell. The second output port is directed to an HCN cell, with a majority of the remaining light sent to fast photodiode where the DCS data is collected while a portion (10%) of the light is sent to an OSA. (b) The DCS spectrum is compared to an absorption taken using a broadband light source. (c) The output of a battery-powered DFC device with the inset showing the power supply.

  We have shown a diode-laser frequency comb source that is a viable platform from which to launch future portable, comb-based spectroscopy, frequency metrology, time-transfer, or ranging applications. To be useful in such applications, chip-scale frequency comb generators must be 1) compact, 2) efficient, 3) self-coherent, 4) tunable, and 5) coherent with other such devices. We have shown that combs can be generated in diode lasers grown commercially, with two-dozen or so combs fitting on each device. We have shown they are efficient enough to be powered by standard AA batteries. Our XCDC measurements show that DFCs output coherent, frequency modulated combs. We have measured the phase profile of the field in both the time and frequency domains, showing that the output is dominated by linear chirp, as expected.[22] Furthermore, their output is tunable over multiple free-spectral ranges with both current and temperature, demonstrating that they are a potential source of stabilizable frequency comb spectra for precision time-transfer or frequency metrology applications. Finally, we demonstrated that our combs are coherent with other devices on the same chip by conducting dual-comb spectroscopy of HCN vapor.

  DFCs fulfill the requirements necessary for ubiquitous and portable application. Though we did not stabilize their spectra carefully in this work, future work could focus on simple injection locking schemes to increase control over the comb spectra and decrease tooth linewidths. As is, teeth are roughly on the order of 100 kHz wide.[20] The comb generation, however, is incredibly simple and was sufficiently stable to enable us to characterize the coherence and operation of these light sources. These results are not an advance in the state-of-the-art for DCS, rather they provide a route to making DCS, and combs more broadly, ubiquitous, with even greater potential impact.

**Acknowledgments:**

We thank Prem Kumar for his early encouragement of this project and David Burghoff for helpful discussions on computational correction of comb spectra. The DARPA SCOUT program provided funding for this project through ARO grant W911NF-15-1-0625.

**Data Availability:**

The data that support the findings of this study are available from the corresponding author upon reasonable request.